\documentstyle[12pt]{article}
\begin{document}
\thispagestyle{empty}

\begin{center}
\LARGE \tt \bf{Non-Riemannian Relativistic Superfluid Hydrodynamics}
\end{center}

\vspace{2.5cm}

\begin{center} {\large L.C. Garcia de Andrade \footnote{Departamento de
F\'{\i}sica Te\'{o}rica - Instituto de F\'{\i}sica - Universidade do Estado do Rio de Janeiro- UERJ

Rua S\~{a}o Fco. Xavier 524, Rio de Janeiro, RJ

Maracan\~{a}, CEP:20550-003 , Brasil.

E-mail : garcia@dft.if.uerj.be}}
\end{center}

\vspace{2.0cm}

\begin{abstract}
Relativistic Riemannian superfluid hydrodynamics used in general relativity to investigate superfluids in pulsars is extended to non-Riemannian background spacetime endowed with Cartan torsion. From the Gross-Pitaeviskii (GP) it is shown that in the weak field Cartan torsion approximation, the torsion vector is orthogonal to the superfluid plane wave velocity. Torsion vector is also shown to be aligned along the vortex direction in the superfluid. The background torsion is shown to induce rotation on the fluid as happens with the acoustic torsion in the analogue non-Riemannian non-relativistic superfluid models. The torsion part of the current would be connected to the normal part of the superfluid velocity while the Riemannian part of the velocity would be connected to the superfluid velocity itself. Magnus effect and the rotation of the superfluid are analysed. Since the Kalb-Ramond field is easily associated with torsion our method seems to be equivalent to the vortex-cosmic string relativistic superfluid method developed by Carter and Langlois to investigate rotating neutron stars.
\end{abstract}

\vspace{1.0cm}

\begin{center}
\large{PACS number(s) : 04.20,02.40h}
\end{center}

\newpage

\pagestyle{myheadings}
\markright{\underline{Non-Riemannian Relativistic Superfluid Hydrodynamics}}

\section{Introduction}
Recently one of the authors has shown  \cite{1,2} that the analog gravitational models in condensed matter physics and fluid flows can support non-Riemannian geometry when there is bulk vorticity in the fluid or superfluid. Actually more recent the same author has shown \cite{3} that, as far as we consider classical fluids with vorticity, the geometrical structure in addition to the acoustic metric \cite{4} in the gauge invariant equations of the fluid discovered by Bergliaffa et al \cite{5} is exactly the acoustic torsion \cite{1}. The importance of extending this result to relativistic fluids \cite{6} is moreless evident, since in neutron stars (pulsars) , although neutrons compressed in star superfluid behave like bosons, individually they are still fermions, and torsion \cite{5} relativistically couples to fermions. Bilic has recently \cite{7} considered relativistic analogue models in Bose-Einstein condensates (BEC). With this physical motivation we consider the splitting of the complex nonlinear Schr\"{o}dinger equation (GP) \cite{8}, in Riemann-Cartan spacetime (RC) , in its real and imaginary parts and from them we compute the current conservation in terms of torsion, which allows us to deduce that in the weak field approximation for torsion, it is orthogonal to the superfluid velocity and that this torsion background induces rotation in the superfluid, much in the same way that non-relativistic superfluids have rotational induced effects from the acoustic torsion \cite{1,3}.  The Letter is organized as follows : In section 2 we write the generalised GP complex equation in the non-Riemannian background spacetime and drawn physical conclusions from the model. In section 3 we compute the  Magnus effect and superfluid rotation for the superfluid in RC background. Section 4 addresses the conclusions and discussions.

\section{Superfluid hydrodynamics in RC spacetime}
Let us now consider the GP equation for superfluid hydrodynamics in Riemannian  spacetime \cite{8} to investigated gravitationally repulsive domain wall solutions of Einstein's equations of general relativity. The GP Riemannian relativistic equation is given by
\begin{equation}
{\Box}{\psi}+F(|{\psi}|^{2}){\psi}=0
\label{1}
\end{equation}
where ${\psi}$ is the scalar wave equation of the condensate ,and ${\Box}={\nabla}^{i}{\nabla}_{i}$ is the Riemannian D'Lambertian wave operator 
\begin{equation}
{\Box}=\frac{1}{\sqrt{-g}}{\partial}_{i}(\sqrt{-g}g^{ik}{\partial}_{k})
\label{2}
\end{equation}
Here $(i,j=0,1,2,3)$. To extend the Riemannian GP relativistic equation ro its non-Riemannian couterpart we simply addopt the minimal coupling in the Riemannian covariant derivative ${\nabla}_{i}$ to the non-Riemannian covariant derivative as
\begin{equation}
D_{i}P^{i} = {\nabla}_{i}P^{i}+ K^{i}P_{i}
\label{3}
\end{equation}
where $K^{i}$ is the trace of Cartan contortion tensor and $P_{i}:= {\partial}_{i}{\psi}$. Taking the order parameter of the condensate as ${\psi}={\nu}e^{i{\alpha}}$ where ${\alpha}$ is the phase of the scalar wave function ${\psi}\bar{\psi}={\nu}^{2}$ where the bar over ${\psi}$ denotes complex conjugation. Substitution of expression (\ref{3}) into equation (\ref{1}) yields
\begin{equation}
{\Box}{\psi}+ K^{i}{\partial}_{i}{\psi} + F(|{\psi}|^{2}){\psi}=0
\label{4}
\end{equation}
Splitting of the GP equation (\ref{4}) into its real and imaginary parts we write the equations in the Madelung fluid form
\begin{equation}
{\partial}_{i}{\alpha}{\partial}^{i}{\alpha}=k^{2}={\alpha}K^{i}{\partial}_{i}ln{\nu}+F({\nu}^{2})+ \frac{{\nabla}^{i}{\nabla}_{i}{\nu}}{\nu}
\label{5}
\end{equation}
and
\begin{equation}
D_{i}j^{i}=0
\label{6}
\end{equation}
where the current $j^{i}$ is given by
\begin{equation}
j_{i}={\nu}^{2}{\partial}_{i}{\alpha}+{\nu}K_{i}{\alpha}
\label{7}
\end{equation}
Note that the conservation equation (\ref{6}) is covariant in the RC spacetime. It is interesting to note that in (\ref{7}) the torsion part of the current could be interpreted as the normal current of the superfluid connected to the velocity $v_{n}$ of the normal part of the fluid, while the potential current would be aassociated with the superfluid velocity of the superfluid phase $v_{s}$. To simplify matters we have consider in this first application that the Cartan contortion is constant in the gauge where $K^{i}=(K^{0}=0,\vec{K})$. The current $J^{i}$ can be expressed in terms of the wave function ${\psi}$ is
\begin{equation}
j^{i}= i[\frac{1}{2}({\partial}^{i}{\psi}\bar{\psi}-{\psi}{\partial}^{i}\bar{\psi})-ln(\frac{{\bar{\psi}}}{\psi})K^{i}]
\label{8}
\end{equation}
An important feature of this model is the definition of the current in terms of the superfluid velocity
\begin{equation}
j^{i}=nv^{i}
\label {9}
\end{equation}
where $n=k{\nu}^{2}$ and the constraint $v^{i}v_{i}=1$ yields 
\begin{equation}
v^{i}=\frac{{\partial}^{i}{\alpha}}{k}+\frac{{{\alpha}K^{i}}}{{\nu}{k^{2}}}
\label{10}
\end{equation}
This expressions immeadiatly tells us that the superfluid velocity is not potential anymore even if k is constant. Here $v^{i}=(v^{0},\vec{v})$. Assuming that k is constant we see that the phase ${\alpha}$ may represent a plane wave function ${\psi}$. Of course in this approximation we consider that the superfluid would be in the laboratory and the metric would be reduced to the Minkowski flat metric  and the superfluid could be consider as on a Minkowski plus torsion non-Riemannian background. This assumption is actually still more compatible with the weak approximation we consider here as ${\vec{K}}^{2}=0$ and $\vec{K}.\vec{\nabla}{\alpha}={\vec{K}}.\vec{v}=0$ which shows that the contortion vector is aligned along an  axis orthogonal to the fluid velocity. In the case of vortex filaments \cite{9}, for example, the torsion vector would be aligned along the axis of vortex filaments. To end up this section we compute the vorticity of the superfluid showing definitely that this relativistic non-Riemannian superfluid model is also nonpotential in the superfluid velocities exactly as in the analogue non-Riemannian acoustic  models considered in the literature. From the general speed above one may easily compute the vorticity by considering the curl of the velocity $v^{i}$ as
\begin{equation}
{\Omega}^{k}=\frac{{\epsilon}^{kmi}K_{i}}{k_{0}{\nu}}{\partial}_{m}[{\alpha}-{\nu}]
\label{11}
\end{equation}
In vector notation the vorticity $\vec{\Omega}$ becomes
\begin{equation}
\vec{\Omega}=-\frac{\vec{K}{\times}{\nabla}[{\alpha}-{\nu}]}{k_{0}{\nu}}
\label{12}
\end{equation}
This expression is similar to the expression obtained for the vorticity in the non-Riemannian geometry of vortex acoustics \cite{1}. The second term in expression (\ref{12}) can be easily written in terms of the condensate wave function ${\psi}$ as
\begin{equation}
{{\Omega}_{1}}^{k}= -\frac{1}{2}\frac{{\epsilon}^{kmi}K_{i}}{k_{0}\sqrt{{\psi}\bar{\psi}}}{\partial}_{m}[ln({\psi}\bar{\psi})]
\label{13}
\end{equation}
where ${\Omega}_{0}$ is the other part of rotation. Let us now compute the example of the flow on a cylinder which is imcompressible. From the above equations we obtain the folllowing equation for  ${\alpha}$
\begin{equation}
{\alpha}(r)=A+\frac{B}{r}
\label{14}
\end{equation}
Therefore the superfluid velocity is 
\begin{equation}
\vec{v}^{irrot}={\nabla}{\alpha}(r)= -\frac{B}{r^{2}}{\vec{e_{r}}}
\label{15}
\end{equation}
and the rotational part of the normal fluid is
\begin{equation}
\vec{v}^{rot}=\frac{{(A+\frac{B}{r})\vec{K}}}{{\nu}{k^{2}}}
\label{16}
\end{equation}
Note that at infinity for a finite and weak torsion the flow speed would vanish.
\section{Magnus effect and superfluid rotation in RC background}
In this section, we consider the non-relativistic Magnus effect \cite{10,11} of the uniform background flow with relative velocity $v^{i}$ in the rest frame of a vortex in the direction of a 3 dimensional unit vector $l^{i}$ gives rise to a force per unit lenght given by the Joukowsky formula  
\begin{equation}
F_{i}={\kappa}{\rho}{\epsilon}_{ijk}l^{j}v^{k}
\label{17}
\end{equation}
wher ${\kappa}$ is the relevant circulation integral and ${\rho}$ is the asymptotically uniform density of the superfluid. Applying this formula to the velocity of the superfluid computed in the last section we may compute the force components due to the  irrotational as well as to the rotation  of the superfluid as
\begin{equation}
{F^{irrot}}_{i}=\frac{\kappa}{k}{\rho}{\epsilon}_{ijk}l^{j}{\partial}^{k}{\alpha}
\label{18}
\end{equation}
\begin{equation}
{F^{rot}}_{i}=\frac{\kappa}{{\nu}k^{2}}{\rho}{\epsilon}_{ijk}l^{j}K^{k}{\alpha}
\label{19}
\end{equation}
Note that this last formula shows us that in the case the torsion vector direction coincides with the $l^{i}$ direction such as $K^{k}={\beta}l^{k}$, the torsion does not contribute to the Magnus effect, due to totally skew symmetry of the Levi-Civita symbol ${\epsilon}_{ijk}$. Let us now compute the rotation tensor ${\omega}_{ij}$ from the expressions (\ref{13}) as
\begin{equation}
{\omega}_{mi}= -\frac{1}{2}\frac{K_{i}{\partial}_{m}[ln({\psi}\bar{\psi})]}{k_{0}\sqrt{{\psi}\bar{\psi}}}
\label{20}
\end{equation}
From this expressions is easy to note that expression (\ref{19}) satisfies the Poincare closure property for the vorticity two surface given by
\begin{equation}
{\partial}_{[l}{\omega}_{mi]}=0
\label{21}
\end{equation}
This is easily seen from the relation
\begin{equation}
{\partial}_{l}{\omega}_{mi}= -\frac{1}{2k_{0}}\frac{K_{i}{\chi}_{l}{\chi}_{m}}{{\nu}^{3}}
\label{22}
\end{equation}
where ${\chi}_{m}:= {\partial}_{m}{\sqrt{{\psi}\bar{\psi}}}$. 
\section{Conclusions}
In this letter we showed that the relativistic formulation of superfluids in the background of spacetimes endowed with torsion leads naturally , from the conservation equation to the rotation of the normal part of the fluid while the potential , metric part of the superfluid remains irrotational in the RC spacetime background. This model leads naturally to the investigation of superfluid neutron stars in backgrounds other than general relativity. An interesting and natural extension of the work discussed in this letter would be to consider the extension of the GP system of equation for a spinor  BEC \cite{12} or a fermionic superfluid such as $^{3}He$ in a non-Riemannian acoustic spacetime. This would be even more interesting since relativistic torsion couples more  naturally with torsion. This work is now in progress. The model presented here is also  similar to the Kalb-Ramond (KR) coupled vortex fibration model for the Relativistic superfluid dynamics by Carter and Laglois \cite{13} to investigate the rotating superfluid neutron stars. The reason is that the KR field is simply associated with torsion in cosmic strings theory which is used by those authors to build their model. It is important to call the attention that a type of Magnus field force is also built in their model \cite{13}.
\section*{Acknowledgement}
I am very much indebt to Prof. P.S.Letelier for helpful discussions on the subject of this paper. Financial support from UERJ (FAPERJ) is gratefully acknowledged. Useful discussions with Milena Siqueira is gratefully acknowledged. 

\end{document}